\RequirePackage{fix-cm}
\documentclass[smallextended]{svjour3}                 
\usepackage[colorlinks=true,urlcolor=red,linkcolor=blue,citecolor=blue]{hyperref}
\usepackage{natbib}
\bibpunct{[}{]}{,}{a}{}{,}
\smartqed  
\usepackage{graphicx}
\journalname{Experimental Astronomy}
\usepackage{array}
\begin{document}

\title{VHE Gamma-ray Observation of Crab Nebula with HAGAR Telescope Array}

\titlerunning{Observation of Crab nebula with HAGAR telescope Array}

\author{B.B. Singh\footnote{Corresponding author \email{bbsingh@tifr.res.in}}\and R.J. Britto \and V.R. Chitnis \and A. Shukla \and L. Saha \and A. Sinha \and B.S. Acharya \and P.R. Vishwanath \and G.C. Anupama \and P. Bhattacharjee \and K.S. Gothe \and B.K. Nagesh \and T.P. Prabhu \and S.K Rao \and R. Srinivasan \and S.S. Upadhya}

\institute{B.B. Singh \and R.J. Britto \and V.R. Chitnis \and A. Sinha \and B.S. Acharya \and K.S. Gothe \and B.K. Nagesh \and S.K Rao \and S.S. Upadhya \at Tata Institute of Fundamental Research, Homi Bhabha Road, Colaba, Mumbai 400005, India 
\and	
A. Shukla \and P.R. Vishwanath \and G.C. Anupama \and T.P. Prabhu \and R. Srinivasan \at Indian Institute of Astrophysics, II Block, Koramangala, Bangalore 560034, India
\and
L. Saha \and P. Bhattacharjee \at Saha Institute of Nuclear Physics, 1/AF, Bidhannagar, Kolkata 700064, India 
\and
R.J. Britto \at Department of Physics, University of the Free State, PO Box 339, Bloemfontein 9300, South Africa  \and
A. Shukla \at Institute for Theoretical Physics and Astrophysics, Universit\"{a}t W\"{u}rzburg, 97074 W\"{u}rzburg, Germany  \and
L. Saha  \at Universidad de Complutense, E-28040 Madrid, Spain  \and
A. Sinha \at Astroparticule et Cosmologie/CNRS/ Université Paris Diderot, Paris-75013 
}
\authorrunning{B.B. Singh et al.} 
\date{Received: date / Accepted: date}
\maketitle

\begin{abstract}
HAGAR is a system of seven Non-imaging Atmospheric Cherenkov Telescopes
located at Hanle in the Ladakh region of the Indian Himalayas at an altitude of
4270 meters {\it amsl}. Since 2008, we have observed the Crab Nebula to assess
 the performance of the HAGAR telescopes. We describe the analysis technique for
 the estimation of $\gamma$-ray signal amidst cosmic ray background. The consolidated
 results spanning nine years of the Crab nebula observations show long term performance
 of the HAGAR telescopes. Based on about 219 hours of data, we report the
 detection of $\gamma$-rays
 from the Crab Nebula at a significance level of about 20$\sigma$,
 corresponding to a time averaged flux of (1.64$\pm$0.09) $\times10^{-10}$
 photons cm$^{-2}$ sec$^{-1}$ above 230 GeV. Also, we perform a detailed study
 of possible systematic effects in our analysis method on data taken with the
 HAGAR telescopes.
\end{abstract}

\section{Introduction}
Crab Nebula is the first source detected in very high energy (VHE)
 $\gamma$-rays \citep{1989ApJ...342..379W} and extensively studied object
 by ground based atmospheric Cherenkov detectors. After the first light
 from the supernova (SN 1054) which was recorded in 1054 AD, it is one
 of the best studied non-thermal
 celestial objects in almost all energy bands of the electromagnetic spectrum.
 The Crab Nebula lies $\approx$ 2 kpc \citep{1968AJ.....73..535T} from the Earth
 at a right ascension (RA) of $05^{h}34^{m}31.97^{s}$ and at a declination
 (DEC) of $+22^{d}00^{m}52.1^{s}$, in the constellation of Taurus
 (J2000 epoch). The Nebula has a diameter of 6 ly, and is expanding at a rate
 of about 1,500 kilometers per second. It is also a nearby pulsar wind Nebula
 (PWN) and the Crab is powered by a 33 ms pulsar that injects relativistic electrons
 into the Nebula. Synchrotron radiation by the relativistic charged particles
 (e$^{\pm}$) results in the emission of radiation from radio to GeV $\gamma$-rays,
 while higher energy (GeV to TeV) gamma rays are thought to result from the Synchrotron
 Self-Compton (SSC) process, ie inverse-Compton interaction of the high energy
 e$^{\pm}$ with the synchrotron photons emitted by themselves
 \citep{2006ARA&A..44...17G, 1992ApJ...396..161D, 1996ApJ...457..253D,
 1996MNRAS.278..525A, 1998ApJ...503..744H}.

The Crab Nebula is considered as the “standard candle” in VHE $\gamma$-ray
 astronomy due to its strong and steady emission. In the past few years,
 discovery of variable $\gamma$-ray emission from the Crab Nebula has been
 reported by the AGILE \citep{2011Sci...331..736T} and
 {\it Fermi}-LAT \citep{2011Sci...331..739A}
 telescopes during
 September 2010. During this period Crab Nebula flux in 100 MeV to 1 GeV energy
 band increased by an order of magnitude in less than a day. However, in the TeV
 energy band, MAGIC and VERITAS telescopes did not see any enhancement in the
 flux during this period \citep{2010ATel.2967....1M, 2010ATel.2968....1O}. The ARGO-YBJ
 collaboration have reported enhanced $\gamma$-ray signals with a median energy
of 1 TeV from the direction of the Crab Nebula, which is consistent with the
 flares detected by AGILE and {\it Fermi}-LAT but the increase in flux is below 5$\sigma$ level
 \citep{2010ATel.2921....1A, 2012ATel.4258....1B, 2013arXiv1307.7041V}.
 Another episode of enhanced emission in MeV-GeV energy band, lasting for
 almost two weeks, took place in March 2013. During this
 period, {\it Fermi}-LAT detected a 20-fold increase in the flux of $\gamma$-rays for
 energies  above 100 MeV. Again, VHE observations carried by VERITAS during this
 period did not show any evidence for increase in $\gamma$-ray flux \citep{2014ApJ...781L..11A}.
 The HAWC \citep{2015ICRC...34..744S} detector also did not find any evidence of variations
 in the Crab flux during the period June 13, 2013 to July 9, 2014. In some of
 the earlier experiments, also there were reports of rare detection of $\gamma$-rays
 from the direction of Crab Nebula, presumably due to enhanced flux above the detection
 threshold \citep{1986Natur.319..127B, 1990Natur.347..364A, 1992A&A...258..412A, 1978ApJ...221..268G}.
 Thus, though occasionally variability in the flux of $\gamma$-ray was reported in
 various energy bands, it appears that Crab Nebula is a steady
 source in TeV energy band \citep{1977ApJ...216..560T, 2005SPIE.5898...22K}. Therefore
 it can still be used
 as a ``standard candle'' source for the calibration of ground based
 atmospheric Cherenkov telescopes.

 HAGAR is an array of non-imaging atmospheric Cherenkov telescopes, began its
 VHE observation of Crab Nebula and other astrophysical sources in the 2008
 observing season. HAGAR has successfully detected flares from blazars
 like Mkn421 \citep{2012A&A...541A.140S} and and Mkn501 \citep{2015ApJ...798....2S}.
 This paper discusses the method used in our search for
 steady $\gamma$-rays using wavefront sampling technique. In the non-imaging
 technique where $\gamma$ like events can not be directly distinguished
 from the cosmic ray background (hadron generated) events. The subtraction of cosmic
 ray background remain always a challenge for the estimation of absolute $\gamma$-ray flux.
 The second most important point is related with observational method, when
 the sky brightness of $\gamma$-ray source and corresponding background regions
 are different. In order to address the challenges associated with this technique
 the other data sets which include observations of fictitious sources
 (dark region of the sky and bright sky region) with special interest to
 the Crab Nebula region have been studied and discussed in detail. These  data sets
 were used to test and validate the method used in subtraction of cosmic ray background
 from $\gamma$-ray source region.

\section{HAGAR}
\begin{figure}
\centering
\includegraphics[width=2.0in,height=2.4in]{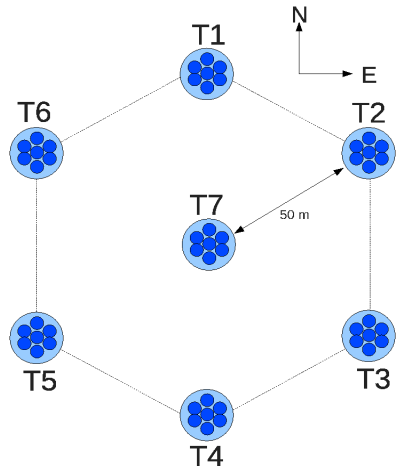}
\caption{Schematic diagram of HAGAR telescope array}
\label{fig1}
\end{figure}

The High Altitude GAmma Ray (HAGAR) observatory consists of an array of seven
 atmospheric Cherenkov telescopes located at the center and corners of a hexagon
 inscribed in a circle of 50 meter radius, which is shown in figure-\ref{fig1}.
 Each telescope consists of seven parabolic glass mirrors of 10 mm thickness and
 0.9 m diameter having f/d=1. These mirrors are front coated and average
 reflectivity in the visible range is around 80$\%$. Each mirror has a UV-blue
 sensitive XP2268B (Photonis) Photo-multiplier tube (PMT), mounted at its focal
 point with 3$^{\circ}$ field of view (FOV) angular mask. The total reflector
 area of all seven telescopes is about 31 m$^{2}$. These telescopes, which are
 based on Alt-Azimuth mounting, are controlled remotely through GUI/Linux based
 system using 17-bit rotary encoders, stepper motors, Microcontroller-based
 Motion Control Interface Units (MCIU) etc. The control system allows to
 achieve a steady state pointing accuracy of 10 arcsec with a maximum slew rate
 of 30$^{\circ}$ per minute for each axis and continuous monitoring of the
 telescope positions. Guide telescopes fitted paraxial to telescope mirrors are
 used to arrive at a pointing model for each telescope. The co-planarity of all
 7 mirrors of a given telescope with its axis is achieved by a series of bright
 star scans. The over-all accuracy in pointing of the mirrorsis about 12 arc
 minutes. Details of the telescope control and the pointing
 model of HAGAR array have been extensively discussed
 \citep{2013ExA....35..489G}. The high voltages to PMTs are controlled and
 monitored through CAEN universal multi-channel power supply system. The analog
 PMT signals are transmitted to the control room located at the center of the
 array (below Tel \#7) through coaxial cables. The pulses of seven PMTs of a
 telescope are linearly added to generate a telescope pulse. A CAMAC and VME
 based data acquisition (DAQ) system has been used for processing of signals
 from individual PMTs as well as telescope pulses. A DAQ consisting of eight
 channel Flash ADC (fast waveform digitizer, Acqiris make) system has been
 used to digitize 7-telescope pulses. A trigger for the initiation of data
 recording is formed when any four out of seven telescope pulses cross
 the pre-defined discriminator threshold within a coincidence window of 60 ns.
 In this paper we restrict our analysis to the data recorded by the digitizer.
\section{Observations and Data sample}
 We observed Crab Nebula extensively with HAGAR telescopes since its inception.
 The observations time period extended from October 2008 to December 2017. The key
 information, the direction cosine of the shower axis, is derived from the
 relative time of arrival of Cherenkov shower front at telescopes, which is
 recorded through 8 bits Agilent (time resolution = 1ns) waveform digitizer.
 Observations are carried in ON-OFF mode (source followed by its background or
 vice-versa). The source and corresponding background region are observed for typical
 duration of 60 minutes each. The declination
 angle of the cosmic ray background region and the duration of observation are
 kept same as those of source to have same zenith angle range. In order to
 assess the performance and systematic in the analysis method several other
 regions like fixed angle (telescopes parked at some zenith angle) and dark
 region (fictitious source) of the sky were also observed. Log of observation
 duration after selection of ON-OFF run pairs taken on the same night and having
 same zenith angle coverage are given in Table - \ref{obs_log}. All types of
 observations (both source and background) were taken on the same night to minimize
 the effect of sky conditions on PMT counting rates. The night sky condition
 and count rates of individual PMTs were monitored throughout the observations.
\begin{table}
\caption{Observation Log for Crab Nebula and related runs}
\begin{center}
\begin{tabular}{c|c|c} \hline
         & Number of ON/OFF  & Duration (hours)  \\
  Source & run pairs         &                   \\\hline
 Crab Nebula       & 241& 219.0   \\
 Dark region       & 108 & 97.5    \\
 Fixed angle       & 98 & 46.1   \\
 Bright sky region & 26 & 24.8  \\\hline
\end{tabular}
\label{obs_log}
\end{center}
\end{table}
\section{Analysis}
The analysis of the HAGAR data has been performed using in-house developed codes
 in the IDL programming language. In this technique, analysis of the data is based
 on the comparison of cosmic-ray events from a $\gamma-$ray source
 region with similar cosmic-ray background region. The complete analysis method
 is grouped into three sub-sections. These are data reduction, estimation of event arrival
 direction and extraction of $\gamma$-ray signal.
\subsection{Data reduction}
The Acqiris DC271 high speed 8 bit waveform digitizer is configured for 1
 nanosecond sampling period, records waveform of PMT pulses at the rate of
 1 GHz. The vertical scale of digitizer was configured for 2 Volts with an offset set at positive 875
 millivolts. The pre-trigger delay was set for 700 ns and the time stamp for
 the triggered event is recorded up to microsecond using cPCI GPS clock.
\begin{figure}
\centering
\includegraphics[width=3.5in,height=1.5in]{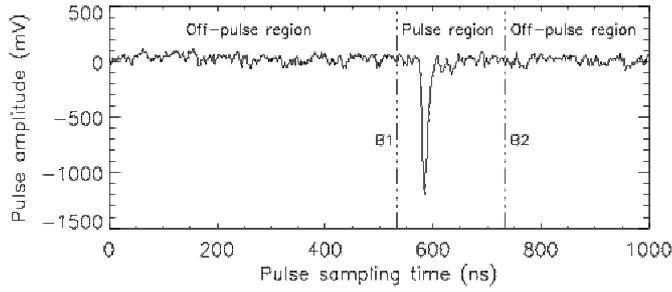}
\caption{Typical Cherenkov (telescope) pulse pattern produced by the flash ADC. The pulses
 of seven PMTs of a telescope are linearly added to generate a telescope pulse. The B1 (50ns)
 and B2 (150ns) indicate pre and post time stamps of Cherenkov pulse region with respect
 to pulse arrival time.}
\label{fig2}
\end{figure}
\begin{figure}
\centering
\includegraphics[width=2.8in,height=1.9in]{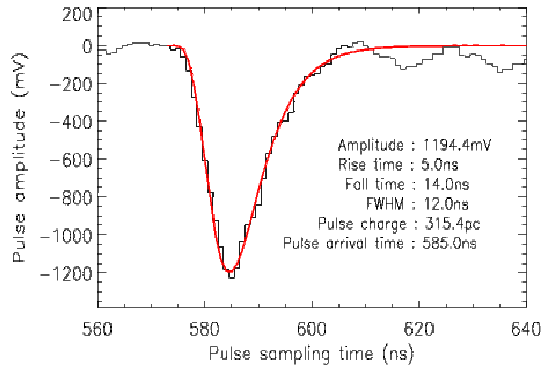}
\caption{Cherenkov pulse fitted with Log-Normal function}
\label{fig3}
\end{figure}
A dynamic window approach was used to locate the Cherenkov pulse region in the
 readout trace of 1000 ns sampling time. A typical waveform produced by the flash
 ADC as shown in the figure-\ref{fig2} indicates pulse and off-pulse regions. The
 Cherenkov pulse window is set for 200 ns and remaining traces of waveform is used
 for the estimation of average night sky background (NSB) during ON-source and
 OFF-source observation. The figure-\ref{fig3} shows Cherenkov pulse
 fitted with Log-Normal function. The pulse parameters such as amplitude, rise
 time, fall time, pulse width (FWHM), pulse charge and pulse arrival time are
 calculated from the fitted pulse shape. The pulse arrival time is defined as the time
 at which the pulse amplitude reaches 95$\%$ of its absolute maximum.
\subsection{Reconstruction of event arrival direction}
 The relative arrival times of telescope pulses are used to reconstruct the
 arrival direction of Cherenkov shower. These relative arrival times are first
 corrected for a fixed time offset called tzero (T0). A finite but constant time
delay (tzero) between telescope channels arises due to the difference in the
 signal path length, propagation delay in processing electronics and transit
 time of PMTs. The tzero values are calculated using data from runs conducted by pointing
 all the telescopes in a fixed direction, such as Zenith, 10$^{\circ}$ North,
 10$^{\circ}$ South etc. For each pair of telescopes, we get an equation of the form
 \citep{2003APh....18..333M}
\begin{equation}
                  \chi^2 = \Sigma w_{ij}(T0_i-T0_j-C_{ij})^2
\end{equation}
where T0$_i$ and T0$_j$ are the tzeros (time offsets), $C_{ij}$ is mean delay respectively
 between a pair of i$^{th}$ and j$^{th}$ telescope and the $w_{ij}$ is weight factor.
 The weight factor
 $w_{ij}=1/\sigma^2_{ij}$ is calculated from $C_{ij}$ distribution. The tzeros values are
 calculated by solving simultaneous equations formed by minimization of
 $\chi^2$ value.
\begin{figure}
\centering
\includegraphics[width=4.75in,height=3.0in]{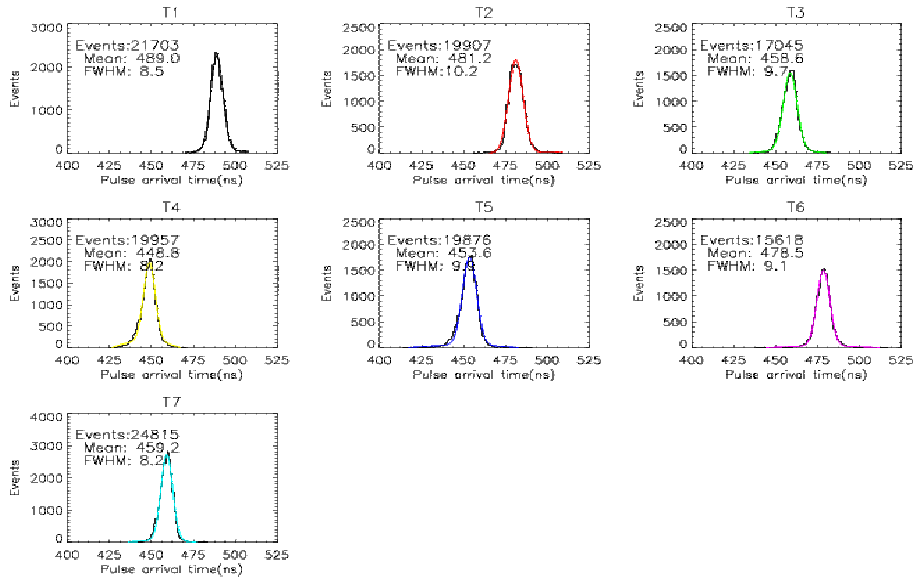}
\caption{Arrival time distribution of telescope T1, T2, T3, T4, T5, T6 and T7
 pulses taken from 10$^{\circ}$ South (Zenith angle=10$^{\circ}$ and Azimuthal
 angle=180$^{\circ}$) run \# 6053}
\label{fig4}
\end{figure}
Figure-\ref{fig4} shows the distributions of pulse arrival time taken from the
 10$^{\circ}$ South direction. The variation in the mean arrival time is due to
 the difference in the geometrical delay arising due to inclination angle,
 relative difference in the z-height of telescopes and tzeros. The average
 time delay between two telescope pulses after correction for geometrical
 delay and z-height from a large sample of data accurately represents the two
 time-offsets.
\begin{figure*}
\centering
\includegraphics[width=4.75in,height=1.8in]{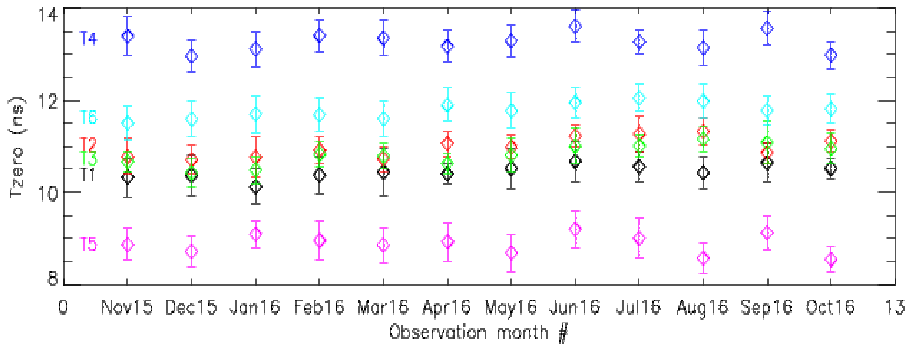}
\caption{Relative time offsets (tzeros) of telescopes T1, T2,...T6 with respect to T7 over period of one year}
\label{fig5}
\end{figure*}
 The data used for estimating the time-offsets of HAGAR telescopes consist of
 only cosmic ray events. Figure-\ref{fig5} shows the time-offsets of telescopes relative
 to the telescope T7 over the period of twelve months. The statistical error in
 the tzeros estimated from different fixed (North, South or Zenith)
 direction events is less than a nanosecond. Any other variation in the tzeros
 of a given telescope could be due to changes in operating (signal path) and
 climatological conditions. Monthly averaged, time-offsets were used in the
 estimation of the arrival direction of events collected in different data sets.

\begin{figure}
\centering
\includegraphics[width=2.8in,height=2.0in]{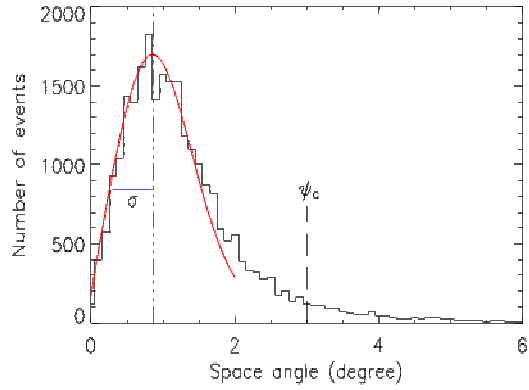}
\caption{Distribution of space angle of events. Data taken from run \# 6053 (10 deg South) and observation duration is 60 minutes}
\label{fig6}
\end{figure}

In the HAGAR array, arrival direction of the shower is estimated using the
 plane front approximation \citep{0954-3899-19-7-016}.
 The arrival direction of the Cherenkov shower can be estimated by minimization
 of $\chi^2$
\begin{equation}
                  \chi^2 = \Sigma w_{i}(lx_i+my_i+nz_i+c(t_i-t_0))^2
\end{equation}
 where $x_i,y_i,z_i$ are the coordinates of the i$^{th}$ telescope, l,m,n the
 direction cosines of the shower axis, t$_i$ the arrival time of the
 showerfront at this telescope and t$_0$ is the arrival time of shower front at
 the origin of the coordinate system. Timing measurement of i$^{th}$ telescope
 is weighted $(w_i=1/\sigma^2_i)$, where $\sigma_i$ is uncertainty in the relative
 timing measurement of Cherenkov shower wavefront which arises due to shower
 fluctuation and arrival direction. The values of l,m,n and t$_0$ are
 calculated by solving equations
 $\partial \chi^2/\partial l=0, \partial \chi^2/\partial m=0,
 \partial \chi^2/\partial t_0=0$ and $l^2+m^2+n^2=1.$ The space angle ($\psi$)
 is an angle between telescope pointing direction and reconstructed direction of the shower and is given by
\begin{equation}
                  cos~\psi = l_{1}.l_{2}+m_{1}.m_{2}+n_{1}.n_{2}
\end{equation}
where ($l_{1},m_{1},n_{1})$ and $(l_{2},m_{2},n_{2}$) are the direction cosines
 of telescope pointing and reconstructed direction of the shower. Figure-\ref{fig6} shows the space
 angle distribution of cosmic ray events for 10$^{\circ}$ South direction run.
 Since opening angle or view cone of HAGAR telescope is 3$^{\circ}$
 cosmic ray showers with incidence angle roughly in the range of $\pm$1.5$^{\circ}$
 with respect to the pointing direction can trigger DAQ.
 The acceptance of showers increases with view cone due to increase in solid angle and
 detection efficiency of showers decreases with zenith angle due to absorption of
 showers. The convolution of
 detection efficiency with incident showers shows maximum around 1 degree. In the
 figure-\ref{fig6} peak of the space angle distribution occurs at
 0.8$\pm$0.1 degree and space angles greater than 3$^{\circ}$ are due to poor fitting of
 shower wavefront.
\begin{figure}
\centering
\includegraphics[width=2.8in,height=2.0in]{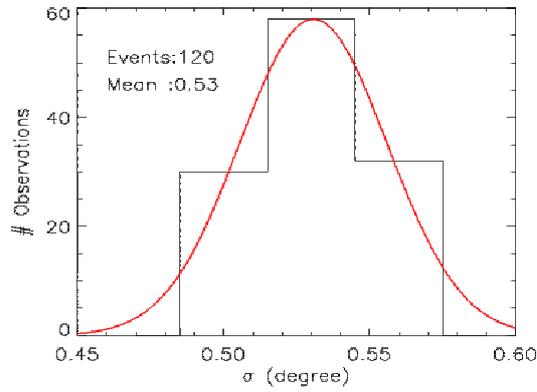}
\caption{Distribution of $\sigma$ of Gaussian function fit to space angle distributions
 for 10$^{\circ}$ North, 10$^{\circ}$ South and Zenith direction runs}
\label{fig7}
\end{figure}
The space angle distribution of events using CORSIKA \citep{2013APh....42...33S}
 almost overlap in full width at half maximum (FWHM)
 and peak position with run data. The standard deviation of a Gaussian fit to
 space angle distribution is obtained and the figure-\ref{fig7} shows the distribution
 of these standard deviations ($\sigma$) obtained from fixed direction runs. The
 mean of the distribution is $0.53\pm0.02$$^{\circ}$.
\subsection{Extraction of $\gamma$-ray signal}
The trigger rate stability is checked as a function of recorded time.
 Any short term contamination in data which arises due to instrument error or
 bad sky condition are rejected by clipping run data. The relative arrival times
 of Cherenkov shower front at telescopes are fitted with a plane and normal to
 this plane gives the direction of arrival of the shower. If the residue
 (observed - expected) delay is greater than 3 ns for any of the telescopes
 then same event is reprocessed after rejecting the telescope having largest
 deviation and this iteration continues till all residues are within the
 3 ns or less than four telescopes are available for the angle fit, whichever
 occurs earlier. In this process almost 10-15$\%$ of events get rejected as we
 need atleast 4 telescopes with valid arrival time of Cherenkov shower front
 for estimating the arrival direction. The ON-OFF runs are paired based on the
 overlapping period (hour angle) taken during the same night.

The excess/deficit $\gamma$-ray signal is estimated by comparison of events
 from source direction to its background direction.
\begin{table*}
\caption{Excess/deficit signal from fictitious source runs}
\centering
\begin{tabular}{c|c|c|c|c} \hline
         &\multicolumn{2}{c|}{Set-I}  & \multicolumn{2}{c}{Set-II}  \\\hline
  Run ID &  5625 & 5626  & 6365  & 6371  \\
   MJD   & 57397 &57397  & 57773 & 57774       \\
RA (HH:MM:SS) & 06:50:29 & 08:05:29 & 06:50:34 & 06:50:34 \\
DEC (DD:MM:SS)& 20:16:41 & 20:16:41 & 22:01:30 & 22:01:30 \\
Total events & 29789 &33191  & 16409 & 19202       \\
Event rate (Hz)& 7.45 & 8.31 & 5.36 & 6.28  \\\hline
Obs. time (minutes)   &\multicolumn{2}{c|}{66.6}  & \multicolumn{2}{c}{51.0}  \\
Rate1 (minute$^{-1}$) &\multicolumn{2}{c|}{-51.1$ \pm$ 3.8}  & \multicolumn{2}{c}{-54.8 $\pm$ 3.7}\\
Constant (C)  &\multicolumn{2}{c|}{0.8984}  & \multicolumn{2}{c}{0.8516} \\
Rate2 (minute$^{-1}$) &\multicolumn{2}{c|}{-0.4$ \pm$ 3.6}  & \multicolumn{2}{c}{1.1$ \pm$ 3.4} \\
\hline
\end{tabular}
\label{fict}
\end{table*}
\begin{figure}
\centering
\includegraphics[width=3.5in,height=2.0in]{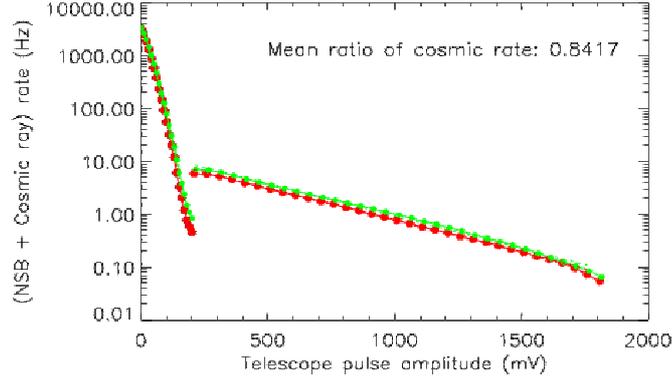}
\caption{Integral event rate of night sky and cosmic Cherenkov light at four-fold
 trigger condition. The red and green points curve correspond to run ID$\#$6365 and
 run ID$\#$6371}
\label{fig8}
\end{figure}
\begin{figure}
\centering
\includegraphics[width=3.0in,height=2.0in]{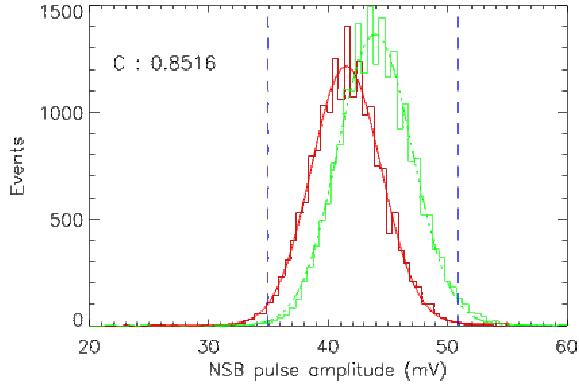}
\caption{Distribution of night sky pulses of a run pair from dark region (data Set-II). The red and green curve correspond to run ID$\#$6365 and run ID$\#$6371}
\label{fig9}
\end{figure}
Data taken with different sky condition or operating condition results in large
 excess/deficit counts. This difference is neutralized through the normalization
 of night sky pulses. The Table-\ref{fict} lists details of two run pairs on
 fictitious source.
 The data Set-I corresponds to a run pair taken on the same night, but different
 region of the sky having the same declination but offset in RA by 75 minutes and
 data Set-II corresponds to a run pair taken on different
 nights but the same region of the sky. The common observation duration for a run pair
 is calculated from the overlapped region of zenith angle range. Figure-\ref{fig8}
 shows a typical integral rate-bias curve in which counting rate (Hz) is plotted as
 a function of the pulse amplitude. The data points (red and green) correspond to the
 data Set-II of the Table-\ref{fict}. The steeply falling curve is due to night sky light
 extracted from the off-pulse region (figure-\ref{fig2}) and flat curve is produced
 from the telescope pulses. The intersection point of the night sky and cosmic ray
 rate-bias curve represent the bias threshold (200 mV) set for the present analysis.
 The mean ratio of ON-OFF cosmic-ray rate deviates from unity because the same operating
 condition could not be maintained during observations.

 These runs were taken from the sky region, which is devoid of any known
 $\gamma$-ray source so the excess/deficit count rates from such run pairs must
 result in a null signal. The run pairs are formed by taking first run as
 an ON-source (fictitious $\gamma$-ray source) and second run as an OFF-source
 (cosmic ray background) without any prior bias in the selection of the sky
 region. The rate (Rate1)
 shows the excess/deficit signal after direct subtraction of OFF-source events from
 corresponding ON-source events. The excess/deficit signal from such source run pair
 deviates from the neutrality condition. Therefore OFF-source events must
 be normalized for the effect which causes a significant difference in the event rate
 of the regular ON-source and OFF-source runs.

 The off-pulse region (figure-\ref{fig2}) can be used to compute a suitable
 normalization factor. Since off-pulse region comprises pulses from the night
 sky light so effect of change in the operating condition of a run pair can be seen
 in the distribution of their night sky pulses. Figure-\ref{fig9} shows the amplitude
 distribution of night sky pulses for run pair of data Set-II. The average
 amplitude of the night sky pulse was calculated for every event. For this purpose,
 pulses of an event for given telescope were averaged in the off-pulse region.
 This average pulse height of a telescope in an event is further averaged over all
 seven telescopes thus yielding a single average pulse per event. The normalization
 constant (C) is given by
\begin{equation}
                 C=\frac{\sum^{l2}_{l1} N^{NSB}_{ON}}{\sum^{l2}_{l1} N^{NSB}_{OFF}}
\end{equation}
The N$^{NSB}_{ON}$ and N$^{NSB}_{OFF}$ are the number of night sky pulses from ON-source
 and OFF-source directions. Two limits ($\it{l1,l2}$) define a common range in
 which only pulses due to night sky light contribute in the distribution.
 The normalization constant is estimated for each selected ON-OFF run pair and then
 the excess/deficit signal is calculated. The sky and operating condition (mostly PMTs high
 voltages) vary from run to run. The excess/deficit signal from an ON-OFF run pair is given by
\begin{equation}
             Signal=\sum^{\psi_c}_{0} N_{ON} - C* \sum^{\psi_c}_0 N_{OFF}
\end{equation}
The $\psi_c$ is the upper limit on the space angle. The N$_{ON}$ and N$_{OFF}$ are
 a number of events ($\psi \le \psi_c$) from the source and background direction and
 constant C is a normalization factor. The excess/deficit rate (Rate2) shows signal
 after the normalization of OFF-source events.
\section{Results}
\subsection{Fictitious source}
The possibility of systematic effects in the excess/deficit signal due to improper
 normalization of the night-sky background levels between the ON and OFF source
 regions of the sky is a major problem for ON-OFF observation technique. The PMT count
 rate and the trigger rates of each ON-OFF pair
 are very sensitive to the night sky conditions. The possibility of any systematic
 error due to the improper normalization of run pair was checked with fictitious
 source observations. Two types of data sets were used to calculate systematic
 errors in the detected signal. The fictitious source observations were taken from
 the different regions of the sky with different brightness. Data sets
 ``Dark region" and ``Fixed angle" described in Table-\ref{obs_log} represent the
 fictitious sources.

\begin{figure}
\centering
\includegraphics[width=1.6in,height=2.05in]{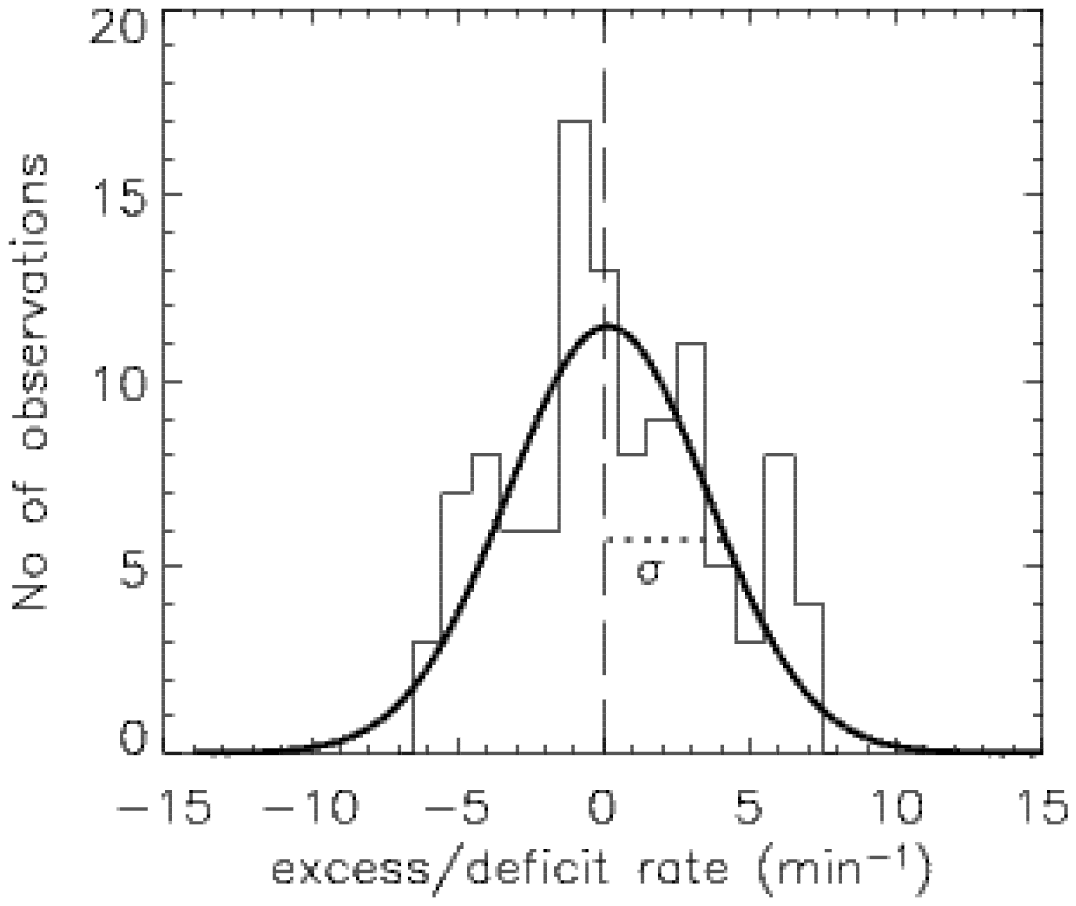}
\includegraphics[width=1.6in,height=2.0in]{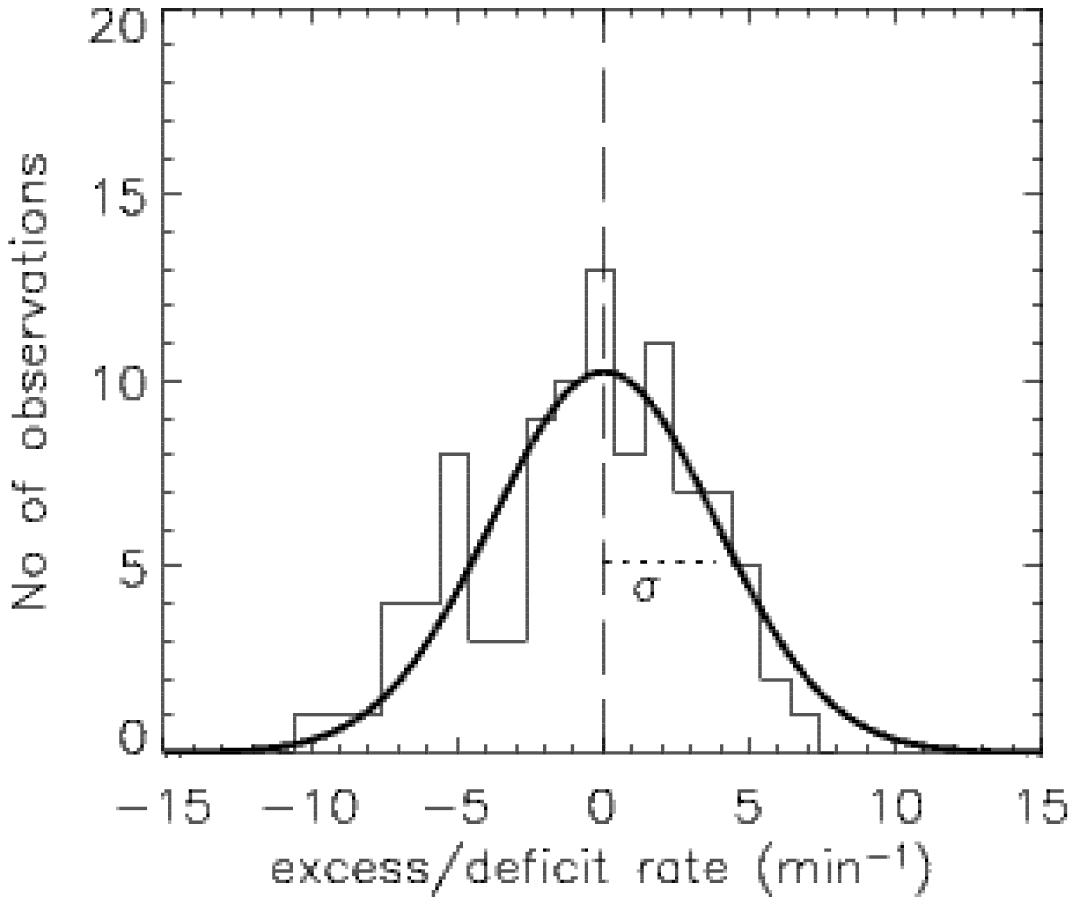}
\caption{Distribution of excess/deficit rates from the run pairs of ``Dark region" (left panel) and ``Fixed angle" (right panel)}
\label{fig10}
\end{figure}

Results of both fictitious source data sets are listed in Table-\ref{fict_src}.
 The figure-\ref{fig10} shows the distribution of $\gamma$-ray signal from the
 fictitious sources. The systematic error in the rate of $\gamma$-ray event per minute
 is about 0.33. In both cases the mean signal is close to zero as expected.
 The run pairs of data set ``Dark region" are
 from the same brightness region of the sky and devoid of any known VHE source,
 therefore the $\gamma$-ray signal derived by selecting any of the run as ON and
 the other as OFF and vice versa would give the same null result. The run pairs of
 data set ``Fixed angle" are from the transiting sky region. These runs have effect of
 varying night sky light, therefore the distribution shown in figure-\ref{fig10} is
 expected to be broader for the ``Fixed angle" runs compared to ``Dark region" runs.
 The statistical error ($\pm$0.37 minute$^{-1}$) and systematic error
 ($\pm$0.33 minute$^{-1}$) of the ``dark region" runs are relevant for the extraction of
 signal from a true source as these runs are analogous to regular $\gamma$-ray source
 observations.
\begin{table}
\caption{Excess/deficit signal:Fictitious source observations}
\begin{center}
\begin{tabular}{c|c|c} \hline
                     & Dark region &  Fixed angle \\\hline
Runs (N)             &  108        &  98          \\
Average Event rate (Hz) & 7.99$\pm1.31$  & 9.49$\pm2.17$  \\
Excess/deficit rate (minute$^{-1}$) &  0.01$\pm0.37$ & -0.02$\pm0.62$ \\
 $\sigma$      &  3.43       & 3.84       \\
Error on mean ($\frac{\sigma}{\sqrt N}$) & 0.33      & 0.39     \\\hline
\end{tabular}
\label{fict_src}
\end{center}
\end{table}
\subsection{Bright sky region}
\begin{figure}
\centering
\includegraphics[width=3.0in,height=2.0in]{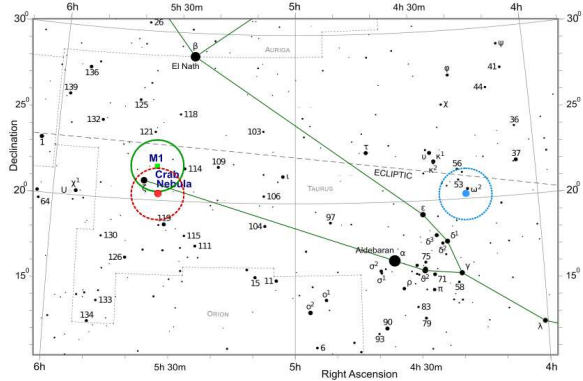}
\caption{Sky map for ``Bright sky region" observations. Green circle (continuous line) shows sky region of Crab
 Nebula observations and red circle (dashed line) and blue circle (dotted line)
 show sky region of fictitious source observations. The angular offset
 between Crab Nebula and fictitious source region is 1.75 degree. Size of
 circle is approximately the FOV of HAGAR telescopes.}
\label{fig10b}
\end{figure}

The FOV of HAGAR telescope is 3 degree and the presence of any bright
star (say magnitude 3 or 4) directly affects the operating conditions and
 triggered events. In a regular observation the high voltages of PMTs are
 adjusted in such a way that 4-fold chance is less than 1$\%$ of trigger rate.
 When the brightness between ON and OFF regions are comparable or slightly
 different then PMT voltages are re-adjusted, mostly by a few volts to maintain
 the chance trigger rate.
 The sky region with the Crab Nebula + $\zeta$ Tauri star (apparent magnitude V=3.010, B-V=-0.164)
 is brighter than the
 corresponding background region. In order to maintain a similar count rates of
 PMTs and low chance rate, the PMTs voltages are re-adjusted, mostly by 50-70
 volts between ON and OFF regions. The effects of change in the operating
 condition between ON-source and
 OFF-source of Crab runs due to the presence of bright star in the ON-source
 region were also checked. The data set ``Bright sky region"
 described in Table-1 and sky map\footnote{https://freestarcharts.com/messier-1} shown in figure-\ref{fig10b} represent
 fictitious source (RA=$05^h35^m36^s$, DEC=$20^d16^m37^s$) and background
 (RA=$04^h20^m36^s$, DEC=$20^d16^m37^s$) regions. Thus, the fictitious source
 region include $\zeta$ Tauri star at the same angular
 offset which is present in regular observations of the Crab Nebula runs and does not include
 Crab Nebula.
\begin{figure}
\centering
\includegraphics[width=3.5in,height=2.0in]{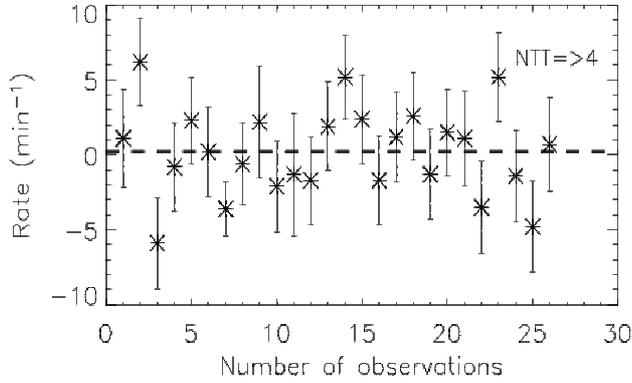}
\caption{Excess/deficit rate per minute from the fictitious source ``Bright sky region" direction}
\label{fig10a}
\end{figure}
These runs with star $\zeta$ Tauri represent as ON-source and corresponding
 pair on the background region taken on same night represent as OFF-source. The average event rate from
 source and background directions are 6.1 $\pm$ 0.3 and 7.3 $\pm$ 0.3 Hz
 respectively. Figure-\ref{fig10a} shows excess/deficit signal as function of
 observations. The average excess/deficit
 count rate from the fictitious source ``Bright sky region" is 0.16 $\pm$ 0.59
 per minute and agree with the null results shown by ``Dark region" and
 ``Fixed angle" data sets.
\subsection{Crab source}
\begin{figure*}
\centering
\includegraphics[width=5.5in,height=1.8in]{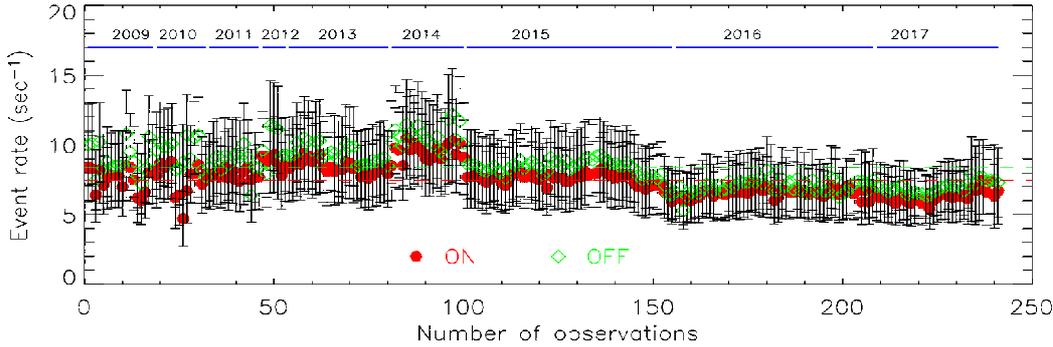}
\caption{Event rate of Crab ON-OFF runs. The blue horizontal line indicates number of observations in the respective
 calendar year}
\label{fig11}
\end{figure*}

\begin{figure*}
\centering
\includegraphics[width=5.5in,height=1.8in]{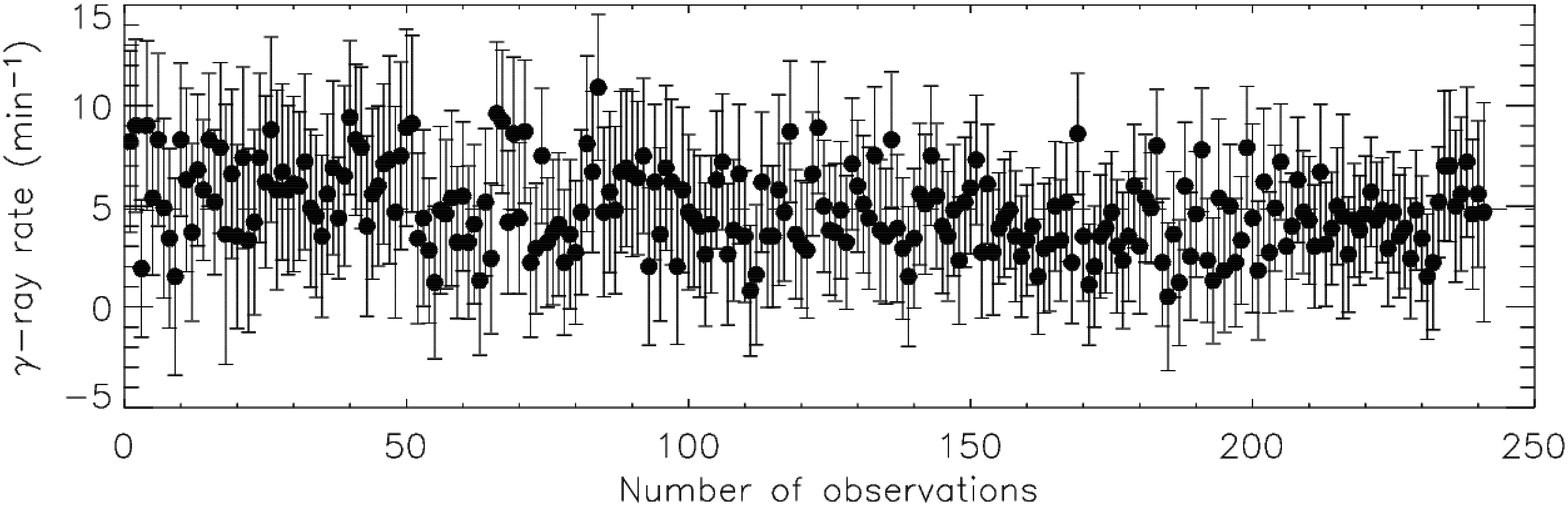}\\
\includegraphics[width=5.5in,height=1.8in]{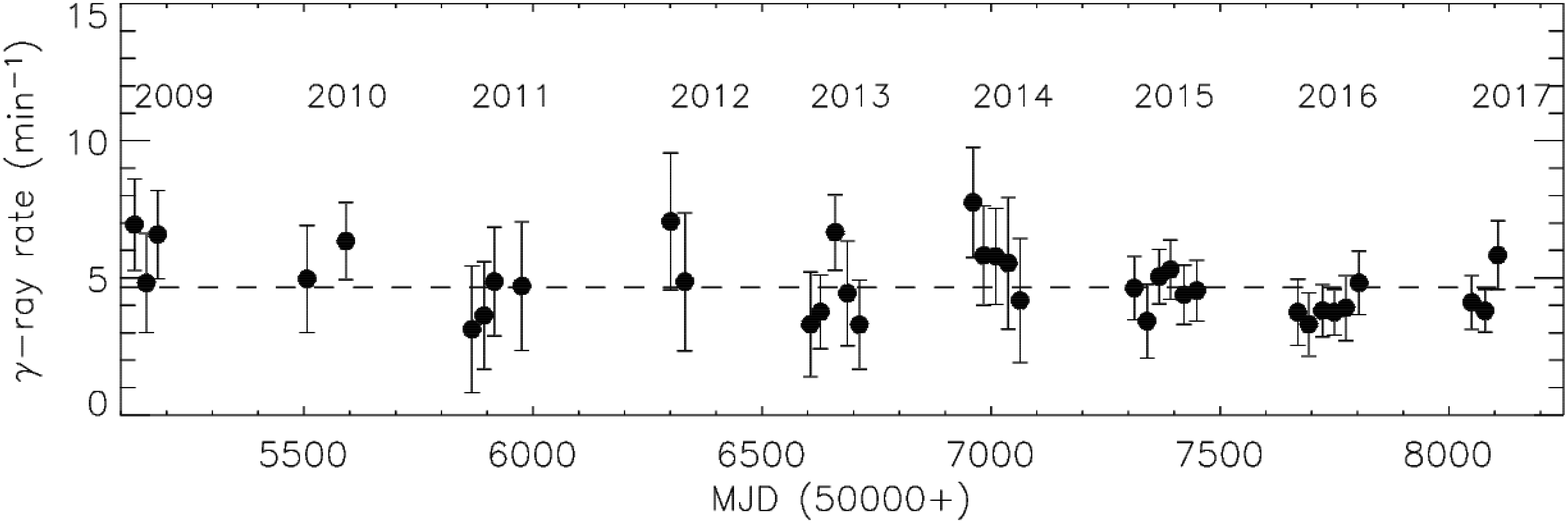}
\caption{$\gamma$-ray event rate from the Crab Nebula as function of time. Upper
 panel: daily light curve, Lower panel: monthly average light curve, The dashed horizontal
 line is the best fit value to a constant $\gamma$-ray rate.}
\label{fig12}
\end{figure*}

\begin{figure}
\centering
\includegraphics[width=2.8in,height=2.0in]{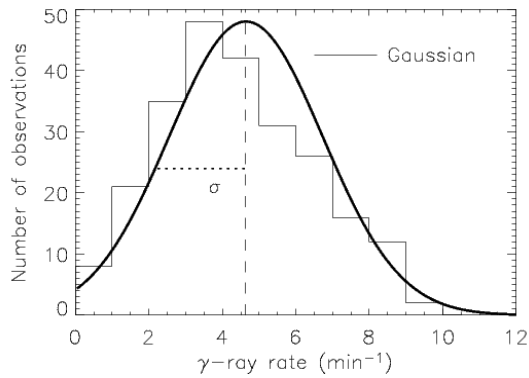}
\caption{Distribution of $\gamma$-ray counts.}
\label{fig12a}
\end{figure}

\begin{figure}
\centering
\includegraphics[width=3.0in,height=2.0in]{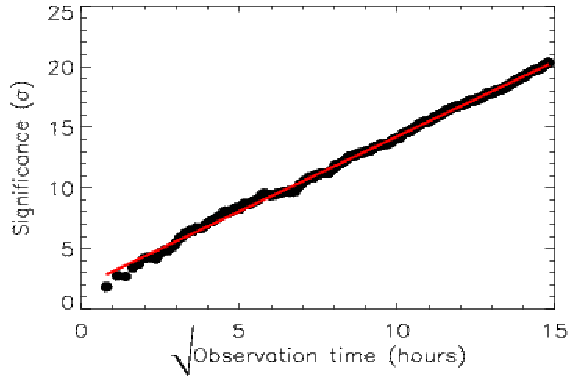}
\caption{Accumulation of $\gamma$-ray signal as a function of observation time}
\label{fig13}
\end{figure}

 The ON-OFF pairs which have common (from zenith/hour angle) duration greater than 30 minutes
 were used in the final analysis. Figure-\ref{fig11} shows event rates of ON-OFF runs
 as a function of the number of observations. Because of slight difference in
 the operating condition (discussed in $\mathsection 5.2$), the event rate of
 OFF-source run is always higher than ON-source run. The space angle
 distribution of ON-source events
 is compared with the corresponding OFF-source events. A cut on the space angle of
 events is imposed to select events from the source direction.
 The space angle limit ($\psi_c$) for accepted events is set at 3$^{\circ}$,
 which is the maximum acceptance angle of HAGAR telescope, and hence $\gamma$-ray signal
 is expected to be within this limit for point sources.
 The normalized background distribution is subtracted from the ON-source distribution
 and excess/deficit of events is calculated. The analysis results of all selected
 ON-OFF run pairs is listed in Table-\ref{crab_src}. The analysis results span over
 nine years of observation data. The column-5 and column-6 show average event
 rates of ON-source and OFF-source runs. The column-7 lists average $\gamma$-ray rate
 detected in each observation year.
 The statistical significance N$_{\sigma}$ was calculated using
 \citep{1983ApJ...272..317L}, is  given by
\begin{equation}    
                 N_{\sigma} = \frac{N_{ON}-C*N_{OFF}}{\sqrt{N_{ON}+C^{2}*N_{OFF}}}  
\end{equation}
where N$_{ON}$ is number of events from the source direction and N$_{OFF}$ are number of
events from the background direction. The column-8 lists total significance $(\sigma)$ detected in
 each observation period and column-9 lists sensitivity of detecting $\gamma$-ray signal.
 Figure-\ref{fig12} shows estimated $\gamma$-ray rate from the Crab Nebula over the
 period of nine years. Each point in the upper panel of the figure shows estimated
 $\gamma$ count rates on daily observation basis and lower panel points show average
 $\gamma$ count rates on a time scale of a month. The low operating voltage of PMTs
 and maximized observation duration during 2015 and 2017 are possible reasons for better stability
 in the monthly averaged signal. Figure-\ref{fig12a} shows the distribution of
 $\gamma$-ray rate per minute from the Crab Nebula. The standard deviation of Gaussian fit
 is used to estimate the error in the signal detection which is due to long term changes
 in the weather and instrument maintenance. This standard deviation is
 1.87 and the error on mean works out to be 0.13 per minute which is used as systematic
 error in the present estimation of $\gamma$-ray rate. We estimate an average
 $\gamma$-ray rate from the Crab Nebula to be $4.64\pm0.23_{sta}\pm0.13_{sys}$ per minute
 at the HAGAR trigger threshold\footnote{Uncertainty in energy threshold is $\pm$12 GeV} of
 230 GeV. Figure-\ref{fig13} shows the significance
 ($\sigma$) of detected signal as a function of observation time. The total statistical
 significance has been 20.3$\sigma$ for 219.1 hours of data and results in a detection
 sensitivity of 1.24$\sigma\sqrt{T}$, where T is the observation time in hours.

 For flux determination, we have calculated the time averaged flux over all the
 HAGAR observations of the Crab Nebula. Energy threshold and collection area applicable
 for average zenith angle of 15$^{\circ}$ was used \citep{2013APh....42...33S}. The average
 flux, thus obtained is ($1.64\pm0.09)\times10^{-10}$ photons cm$^{-2}$ sec$^{-1}$
 for energies above 230 GeV, where the quoted error is only statistical.

 The accuracy of the measured $\gamma$-ray rate and flux depends on the accuracy of the
 reconstruction of the arrival direction and normalization of background cosmic ray events
 in an ON-OFF run pair. Additional uncertainties can arise due to the
 possible variation of the trigger rates as well as due to offsets in the telescope
 pointing, large zenith angles of observations and the applied cuts in the event
 selection process. All these addup to systematic errors.
\begin{table*}
\caption{Excess/deficit signal:Crab Nebula observations}
\centering
\begin{tabular}{c|c|c|c|c|c|c|c|c} \hline
Year& MJD      &Runs&Duration& \multicolumn{2}{c|}{Average Event rate (Hz)} &$\gamma$-rate&Significance&$\frac{\sigma}{\sqrt T}$ \\
\cline{5-6}
    &          &    &(hours) & ON-source & OFF-source&min$^{-1}$     &N$_{\sigma}$      &               \\\hline
2009&55127 - 55188 &18 & 11.2 & 7.54 $\pm$ 0.80 & 8.92 $\pm$ 1.04 & 6.13 $\pm$ 0.98 & 6.26 & 1.63 \\
2010&55500 - 55596 &14 &  9.3 & 7.55 $\pm$ 1.18 & 9.40 $\pm$ 1.02 & 5.83 $\pm$ 1.14 & 5.09 & 1.63 \\
2011&55861 - 55976 &14 &  9.2 & 7.85 $\pm$ 0.39 & 8.75 $\pm$ 0.84 & 4.09 $\pm$ 1.07 & 3.85 & 1.49 \\
2012&56299 - 56332 & 7 &  4.6 & 8.77 $\pm$ 0.53 & 9.97 $\pm$ 1.32 & 6.08 $\pm$ 1.78 & 3.41 & 1.76 \\
2013&56599 - 56714 &27 & 26.6 & 8.41 $\pm$ 0.48 & 9.38 $\pm$ 0.62 & 4.45 $\pm$ 0.71 & 6.28 & 1.52 \\
2014&56956 - 57064 &20 & 19.5 & 9.52 $\pm$ 0.57 &10.66 $\pm$ 0.78 & 5.86 $\pm$ 0.89 & 6.55 & 1.44 \\
2015&57306 - 57456 &55 & 55.7 & 7.54 $\pm$ 0.45 & 8.35 $\pm$ 0.61 & 4.64 $\pm$ 0.46 &10.20 & 1.43 \\
2016&57663 - 57811 &53 & 51.8 & 6.55 $\pm$ 0.31 & 7.03 $\pm$ 0.48 & 3.86 $\pm$ 0.43 & 8.97 & 1.19 \\
2017&58043 - 58112 &33 & 31.1 & 6.74 $\pm$ 1.15 & 7.21 $\pm$ 1.80 & 4.43 $\pm$ 0.56 & 7.91 & 1.37 \\\hline
Average& All data  &241&219.0 & 7.47 $\pm$ 1.08 & 8.35 $\pm$ 1.36 & 4.64 $\pm$ 0.23 &20.30 & 1.24 \\\hline
\end{tabular}
\label{crab_src}
\end{table*}
\begin{figure}
\centering
\hspace*{0.2in}
\includegraphics[width=4.5in,height=3.0in]{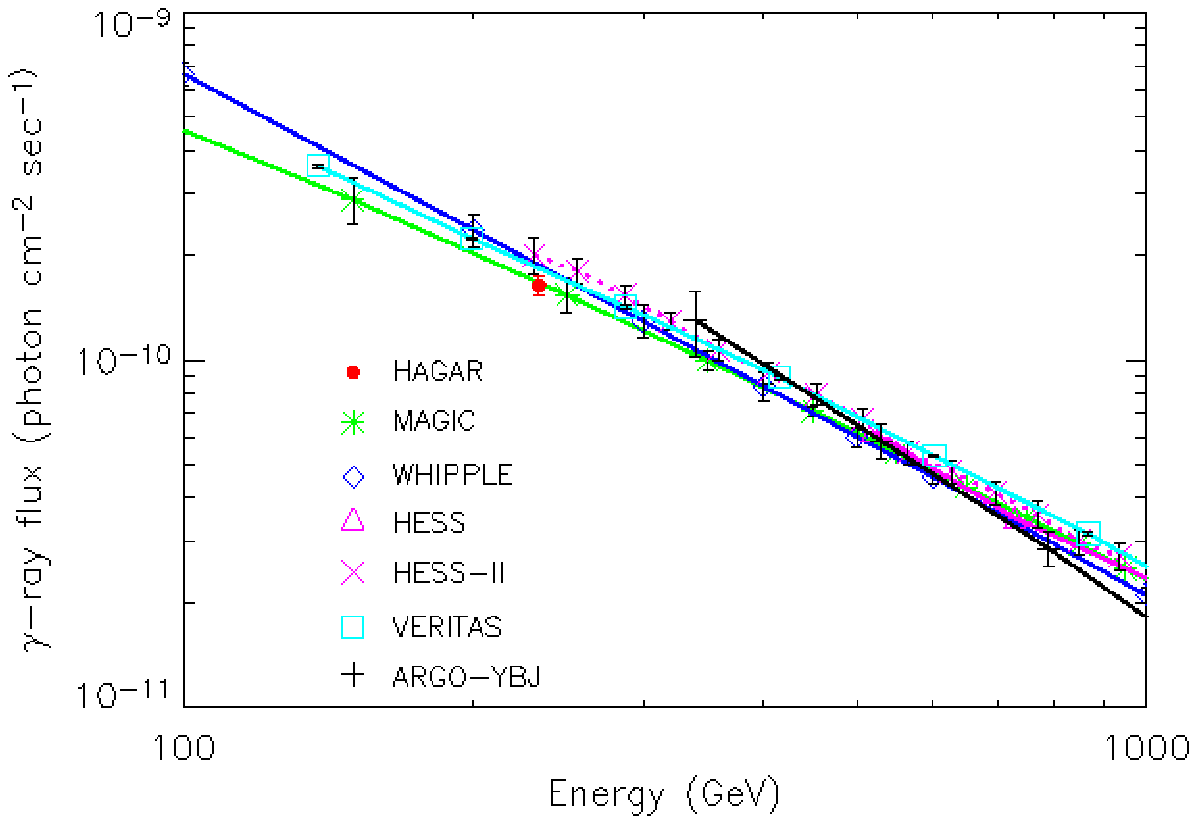}
\caption{Time averaged integral flux detected from the Crab Nebula and compared
 with measurements from other VHE telescopes}
\label{fig14}
\end{figure}
\section{Conclusions}
\begin{enumerate}
\item {The data analysis procedure for the extraction of $\gamma$-ray signal
 using the wavefront sampling HAGAR array has been described in detail.
 We have tested this data analysis method on dark regions devoid of
 any known $\gamma$-ray sources to show that the method does not show
 any fake sources or spurious $\gamma$-ray signal. It is also verified that the
 present normalization method can efficiently equalize cosmic ray events in
 the ON-OFF run pairs.}
\item  {A flux of ($1.64\pm0.09)\times10^{-10}$ photons cm$^{-2}$ sec$^{-1}$ VHE photons
 of energies greater than 230 GeV from the Crab Nebula was
 detected by the HAGAR telescope array at a statistical significance of
 $\sim20\sigma$ over the observation period of 219.1 hours spanning nine years.
 Figure-\ref{fig14} shows the measured flux which is consistent with earlier
 detections by Whipple \citep{1998ApJ...503..744H}, MAGIC \citep{2015JHEAp...5...30A},
 HESS \citep{2006A&A...457..899A, 2015arXiv150902902H}, VERITAS \citep{2015arXiv150806442K} and
 ARGO-YBJ \citep{2015ApJ...798..119B} telescopes.}
\item {Referring to figure-\ref{fig12}, we do not see any significant evidence for the
 variation in the detected signals of $\gamma$-rays from the Crab Nebula during our
 observations spanning nine years.}
\item {Detection of flares from AGNs like Mkn421 \citep{2012A&A...541A.140S} and
 Mkn501 \citep{2015ApJ...798....2S} with the HAGAR telescopes
 has already established its sensitivity to flaring sources. A long term monitoring of
 Mkn421 \citep{2016A&A...591A..83S}
 with HAGAR has been used in multiwave band studies. The blazar 1ES1959$+$650 has also been
 observed during its high active state. The analysis procedure discussed for
 signal extraction uses flash ADC data which is much more robust than CAMAC data used
 in earlier analysis. Improved background subtraction and lower statistical error
 motivate monitoring of sources below 50$\%$ of the Crab flux unit. The HAGAR telescope size and
 FOV are small but advantage of high altitude location has achieved a lower energy threshold. 
 Any alert from a wide field of view ({\it Fermi}-LAT) or IACT telescopes will be source of interest
 for HAGAR telescopes. The follow up or dedicated observations based on such alerts would be main
 targets for future observations.}
 \end{enumerate}
\begin{acknowledgements}
We thank the engineering and technical staff of IIA and TIFR who have taken
 part in the construction, installation, maintenance of telescopes and data
 acquisition setup. Also, we thank the team at the HAGAR site for successful
 observations and data management. We also thank to Kevin Meagher and
 Markus Holler for providing VERITAS and HESS-II measurements of the Crab nebula.  
\end{acknowledgements}
\bibliographystyle{spbasic}
\bibliography{astp_author}

\end{document}